\documentstyle[11pt,paspconf]{article}

\begin{document}

\title{ MPS on the Hunt for Planets}
\author{Sun Hong Rhie}
\affil{Physics Department, University of Notre Dame, IN 46530}

\begin{abstract}
Planetary systems toward the Galactic Bulge can be detected through microlensing 
measurements.  The microlensing planet search technique has some unique merits: 
low-mass planets can be detected from
the ground; the Galactic family of planetary systems can be sampled unbiased;  
the time scale for the completion of each event is relatively short.  
The unambiguity of the underlying low-multiplicity point lenses is an indispensable
element that allows robust interpretaion of the events.  

We have found the first circumbinary planet in the microlensing event
MACHO-97-BLG-41.  We have found evidence of a low mass planet 
(few Earths to Neptune mass) in a very high magnification event MACHO-98-BLG-35.
We emphasize the necessity for coherent 
searches of microlensing planets  with a network of instruments in the 
southern hemisphere.  The network can also be vitally instrumental for follow-up 
observations of  SN1a's and GRB's  which share with microlensing planets 
the transiency as well as the clues to our quest of the origins.  
 
\end{abstract}

Microlensing Planet Search project (MPS http://bustard.phys.nd.edu/MPS/) has  
been monitoring ongoing microlensing events toward the Galactic center alerted by 
microlensing survey projects (MACHO, OGLE, and EROS)  from 1.9m telescope 
at Mt. Stromlo observatory since 1997 in search of planets that may orbit faint 
microlensing stars.   
We monitor photometrically, and the existence of a planet may be 
revealed in a lightcurve that deviates from the family of single lens (bell-shape)
lightcurves. These planetary signals owe to the singular nature of photometric 
microlensing, and they are not necessarily compromised in magnification strength as 
the mass of the planets decreases: they become rarer and briefer, however.  
For earth mass planets, the detection probability is about  $2\%$  
and the typical duration of the planetary signals is expected to be 3-5 hours 
(Bennett and Rhie 1997).  

MACHO-97-BLG-41 (Bennett et al 1999) was the second ``lunatic event" (where a non-single
lens signal appears while the Moon is close to the bulge) of the first season at
MSO that started in May 1997, for which we scrambled taking 30 second exposures to
avoid saturation.   We   continued the coverage of the event in  an expectation of 
a gentle stellar peak whose relative time position and peak value are important 
in determining the possible planetary parameters.  Instead of a mundane gentle peak, 
MACHO-97-BLG-41 was treated with a high magnification plateau (low impact distance) 
that eventually led to stellar caustic crossing.  
This ``stealth bomber" caustic  crossing was greeted with a bit of surprise perhaps 
because  one of the most cited papers, Gould and Loeb (1992), studied giant 
microlensing planets  using an approximation that effected excision of 
the central caustics.   The notion of central caustics was newly emblazoned in the
community, and Griest and Safizadeh (1998) were inspired to examine the effects of
the stellar caustics of high magnification planetary binary lenses.
It is apparent that diagrams, pictures, and mental images 
play a special role (Thorne, Douglas, and MacDonald 1986) in physics even when the 
equations are tractable unlike Einstein equation and the topology in 2d plane not 3+1.
In a planetary binary lensing, the central caustic is rather small, however, and the  
long time span of the feature in the main peak of the MACHO-97-BLG-41 would remain  
unquestioned and unexplained.   Now, we know from an analysis of MPS and MACHO/GMAN
data that the microlens of the MACHO-97-BLG-41 
is a triple lens (Rhie and Bennett 1999)  consisting of a Jovian mass planet orbiting 
a binary star (Bennett et al 1999), and  the central caustic 
is more or less the binary lens central caustic of the binary star.     
Circumbinary planets are another unique territory of microlensing technique
because of its naturally unbiased sampling.   

The experience with the central caustic of the MACHO-97-BLG-41 was handy when
MACHO-98-BLG-35 showed up with peak magnification $\sim 80$.  With our alert
for high probability of  microlensing planet detection,  Microlensing Observation
in Astrophysics collaboration (MOA http://www.phys.vuw.ac.nz
/dept/projects/moa/) embarked on high time resolution imaging from 
New Zealand.  The data nicely complement MPS data where a weak signal of planetary 
deviation was seen.  The combined data show evidence of a low mass planet 
without a giant (Rhie et al 1999b).  It was the highest magnification event so far
out of more than 300 microlensing alert events toward the Galactic center. 
 
We also monitor the events toward the Magellanic Clouds when it deems for 
gathering crucial auxiliary information on the microlenses.  The caustic crossing
binary microlensing event MACHO-98-SMC-1 was one such case
(Rhie et al 1999a and references therein for the global campaign).  The 
global campaign vividly demonstrated an inevitable:  mismatch of (unknown) systematics, 
"evolving data",  and uncertainties in lens parameters polarized between collaborations
(astro-ph/9907247).  It is time to build a ``$2\pi$-detector" with known systematics,
spreaded over South America, Australia (and New Zealand), and South Africa, as was 
recommended by the ExNPS panel (Elachi 1995).     

\acknowledgements  This was supported in part by the NSF and NASA.


\begin{references}
\reference Bennett et al 1999, Nature, submitted, astro-ph/9908038 
\reference Bennett and Rhie 1997, ApJ, 472, 660     
\reference Elachi et al 1995, http://techinfo.jpl.nasa.gov/www/ExNPS/OV.html
\reference Gould and Loeb 1992, ApJ, 396, 104
\reference Griest and Safizadeh 1998, ApJ, 500, 37 
\reference Rhie et al 1999a, ApJ, in print; 1999b, ApJ, submitted, astro-ph/9905151
\reference Rhie and Bennett 1999, in preparation
\reference Thorne, Price, and MacDonald  1986,
   The Black Holes: The Membrane Paradigm, Yale University Press
\end{references}
\end{document}